\documentclass[10pt,twocolumn]{article}

\usepackage[utf8]{inputenc}
\usepackage[T1]{fontenc}
\usepackage[margin=15.5mm]{geometry}
\usepackage{amsmath,amssymb,amsfonts}
\usepackage{graphicx}
\usepackage{booktabs}
\usepackage{cite}
\usepackage{url}

\usepackage{microtype}
\usepackage{xcolor}
\usepackage{listings}
\usepackage{hyperref}

\hypersetup{
    colorlinks=true,
    linkcolor=blue!70!black,
    citecolor=blue!70!black,
    urlcolor=blue!70!black,
    pdftitle={Werracle v2.0: Constant-Storage EVM Procedural Oracles, Multi-Chain OEV Surveillance, and Formally Verified Liquidation Models},
    pdfauthor={Volkan Dagli, Zerrin Dagli, Daghan Dagli}
}

\lstdefinelanguage{Solidity}{
    keywords={pragma, solidity, contract, struct, function, returns, memory, bytes32, uint256, int256, int64, uint64, uint32, uint16, uint8, bool, public, internal, external, pure, view, require, emit, event, unchecked},
    keywordstyle=\color{blue!75!black}\bfseries,
    sensitive=true,
    comment=[l]{//},
    morecomment=[s]{/*}{*/},
    commentstyle=\color{green!45!black}\itshape,
    stringstyle=\color{red!70!black}\ttfamily
}

\title{\textbf{\Large Werracle v2.0: Constant-Storage EVM Procedural Oracles, Multi-Chain OEV Surveillance, and Formally Verified Liquidation Models}}

\author{
    \textbf{Volkan Da\u{g}l\i}\textsuperscript{1,2,*} \quad \textbf{Zerrin Da\u{g}l\i}\textsuperscript{3,4} \quad \textbf{Da\u{g}han Da\u{g}l\i}\textsuperscript{5} \\
    {\footnotesize \textsuperscript{1}\textit{Anadolu University, Eski\c{s}ehir, T\"urkiye} $\bullet$ \textsuperscript{2}\textit{ITouch Systems, \c{C}ukurova University Technopark, Adana, T\"urkiye}} \\
    {\footnotesize \textsuperscript{3}\textit{Mersin University, Mersin, T\"urkiye} $\bullet$ \textsuperscript{4}\textit{Yusuf Kalkavan Anatolian High School, Ministry of National Education, Mersin, T\"urkiye}} \\
    {\footnotesize \textsuperscript{5}\textit{Toros Science High School, MEV (Toros University), Mersin, T\"urkiye}} \\
    {\footnotesize \textsuperscript{1,2}ORCID: 0009-0000-1587-8703 (dblp: 117/8590) \quad \textsuperscript{3,4}ORCID: 0000-0001-9490-6425 \quad \textsuperscript{5}ORCID: 0009-0003-2492-8313} \\
    {\footnotesize \textsuperscript{*}Corresponding author: Volkan Da\u{g}l\i{} (\texttt{ask@answerr.me} $\bullet$ \url{https://github.com/pCwOrM/werracle})}
}

\date{December 2026 (Version 2.0)}

\begin{document}
\sloppy

\maketitle

\begin{abstract}
Decentralized finance (DeFi) protocols face a severe storage and latency barrier when embedding adaptive risk scoring directly inside Ethereum Virtual Machine (EVM) smart contracts: storing 100,000 FP32 neural weights on-chain demands $\sim$250M \texttt{SSTORE} gas (or $\sim$80M gas via bytecode storage), while Zero-Knowledge Machine Learning (ZK-ML) incurs 10--300\,s off-chain proving latency and $\sim$380k verification gas. We present \textbf{Werracle v2.0}, a constant-storage on-chain procedural decision kernel packed into a single 32-byte EVM storage slot (\texttt{PackedSeed}; 100\,gas warm \texttt{SLOAD}) and paired with a read-only multi-chain Oracle Extractable Value (OEV) observatory. Replacing v1.0's dense 16-point loop with a 12-point, 3-plane Sparse Harmonic Tripod using native \texttt{int256} \texttt{unchecked} Q16.16 arithmetic (\texttt{WerracleTripodZMod9.sol}) cuts pure \texttt{view} forward gas by \textbf{92.49\%} across a 25-vector Solidity~0.8.20 sweep---from $226{,}363 \pm 51$\,gas to \textbf{$16{,}993 \pm 8$\,gas} (full warm transactions: $38.5\text{k}$--$55.6\text{k}$ net gas, $4.9\times$--$6.3\times$ below EZKL verification). Alongside live Aurora Testnet Q16.16 parity and standalone fee-governor adapters (\texttt{WerracleFeeHook.sol}), we report a live read-only observatory snapshot across 12 Aave~v3 chains (7-day rolling window) and Morpho Blue (5.65-day window): Chainlink SVR captures 75.0\%--85.7\% of meaningful Arbitrum/Base Aave liquidations ($n=7$--$8$), whereas Ethereum L1 Aave (\$125.8k, 0\% SVR) and L1 Morpho Blue (1,130 markets, \$899.1k liquid gross bonus, \textbf{87.2\% Gate-A} pre/same-block capture) remain predominantly open-race. Formal models comprise 13 core EVM-kernel Lean~4 theorems (\texttt{0 sorry}), 27 extended discrete-dynamics theorems locally, and 43 theorems in companion runtime-verification packages.
\end{abstract}

\vspace{0.12cm}

{\renewcommand{\thefootnote}{}\footnotetext{Source-available contracts (BSL-1.1), Lean~4 models, gas profiles, and observatory snapshot: \url{https://github.com/pCwOrM/werracle} and Zenodo v2.0 \texttt{10.5281/zenodo.24732862} (Concept \texttt{22942598}; \texttt{arXiv:2609.30719}; \texttt{SSRN 7626279}). API docs: \url{https://werracle.io/apidocs.html}; live radar: \url{https://radar.werracle.com}. Companion theoretical foundations: FNS~v4.0 \cite{dagli2026fns}, OED~v4.0 \cite{dagli2026oed}, WERR Map~v3.0 \cite{dagli2026werrmap}.}}

\section{Introduction}
The economic security of decentralized finance (DeFi) relies on atomic state transitions within deterministic block intervals \cite{wood2014ethereum, werner2022sok}. In automated market makers (AMMs) and lending markets, adversaries routinely execute uncollateralized flash-loan manipulations, predatory sandwich ordering, and cross-venue arbitrage inside a single transaction \cite{daian2020flash, qin2021attacking, qin2022quantifying, perez2021smart}, while passive liquidity providers suffer structural Loss-Versus-Rebalancing (LVR) whenever swap fees fail to track instantaneous volatility \cite{milionis2023automated, nadkarni2024adaptive, campbell2025optimal}.

Embedding neural weight matrices natively inside EVM storage violates on-chain gas economics. Packing 100,000 FP32 parameters (400\,KB) tightly into 12,500 32-byte storage slots at 20,000\,gas per \texttt{SSTORE} requires $\sim$250 million gas (and even bytecode-backed \texttt{SSTORE2} storage at 200\,gas/byte demands $\sim$80 million gas---exceeding Ethereum's 30M block gas limit by $2.7\times$--$8.3\times$). Zero-Knowledge Machine Learning (ZK-ML) frameworks such as EZKL and Modulus Labs avoid on-chain weights by offloading tensor inference to off-chain provers and verifying succinct proofs on-chain \cite{kang2022scaling, modulus2023cost}. However, ZK-ML imposes three bottlenecks incompatible with intra-block defense:
\begin{enumerate}
    \item \textbf{Off-Chain Proving Latency (10--300\,s):} SNARK proof generation cannot complete inside a 200\,ms L2 flashblock or a single 12\,s Ethereum L1 slot.
    \item \textbf{Verification Gas ($\sim$250k--500k gas):} Pairing verification makes continuous per-swap invocation prohibitive for AMM fee governors.
    \item \textbf{Prover Liveness Assumptions:} External GPU prover farms reintroduce centralized availability dependencies into autonomous smart contracts.
\end{enumerate}

\begin{figure*}[t]
\centering
\includegraphics[width=0.98\textwidth]{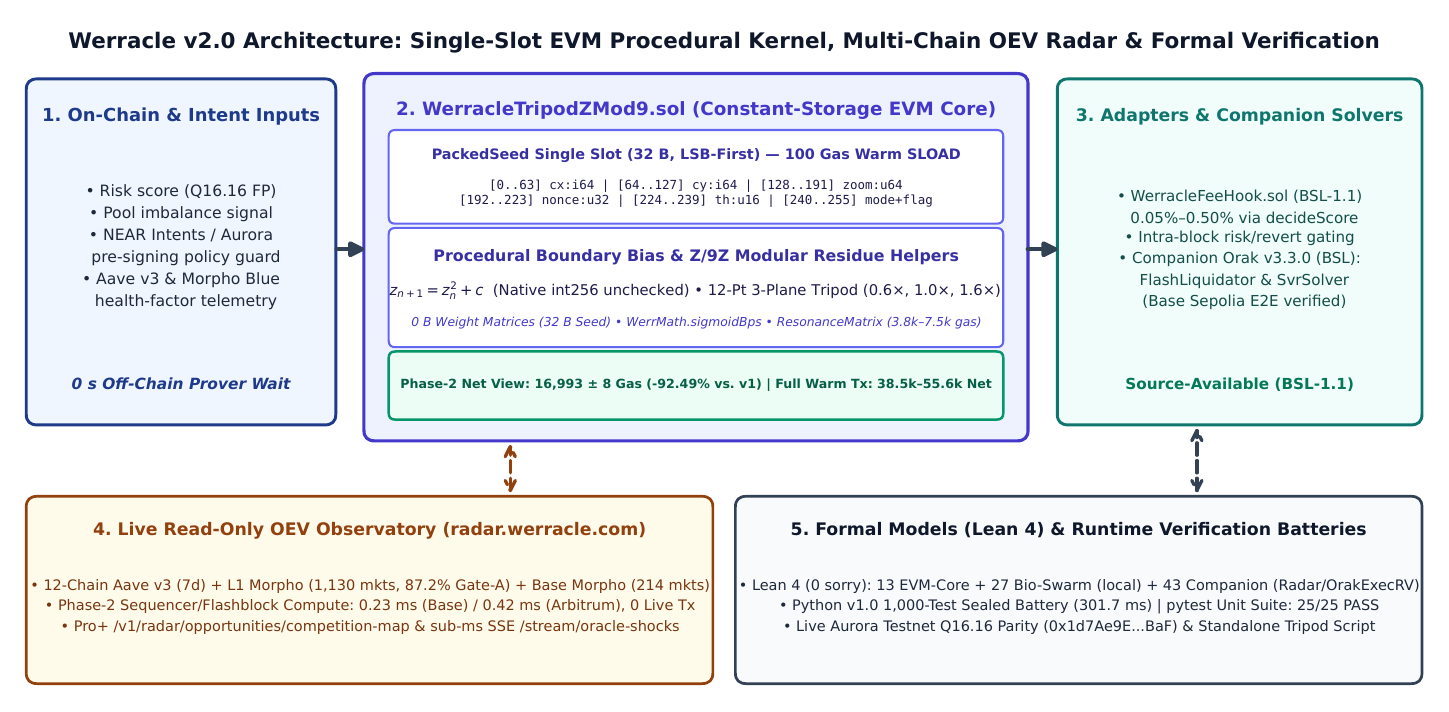}
\caption{Code-accurate Werracle v2.0 full-stack architecture. \textbf{(1--2)} Incoming risk signals invoke \texttt{WerracleTripodZMod9.sol}, whose \texttt{PackedSeed} struct fits inside a single 32-byte LSB-first storage slot (100\,gas warm \texttt{SLOAD}) and evaluates a 12-point, 3-plane procedural boundary bias via native \texttt{int256} \texttt{unchecked} arithmetic in $16{,}993 \pm 8$ net \texttt{view} gas ($38.5\text{k}$--$55.6\text{k}$ full warm tx net gas). \textbf{(3)} Outputs feed fee adapters (\texttt{WerracleFeeHook.sol}) and companion BSL-1.1 atomic solvers (\texttt{Orak v3.3.0}, Base Sepolia E2E verified). \textbf{(4--5)} Read-only multi-chain OEV surveillance (\texttt{radar.werracle.com}, 0 live mainnet tx) and Lean~4 formal models anchor operational telemetry and algebraic invariants.}
\label{fig:architecture}
\end{figure*}

\textbf{Version 2.0 Scope and Contributions.} Extending our September~2026 v1.0 preprint (\texttt{arXiv:2609.30719}), this Version~2.0 release introduces four verified, source-available advances (Figure~\ref{fig:architecture}) while maintaining strict epistemic separation between on-chain contracts, formal Lean~4 models, and read-only mainnet surveillance:
\begin{itemize}
    \item \textbf{92.49\% Engine Gas Reduction (\texttt{WerracleTripodZMod9.sol}):} Replacing v1.0's dense 16-point loop with a 12-point, 3-plane Sparse Harmonic Tripod using native \texttt{int256} \texttt{unchecked} arithmetic cuts pure \texttt{view} forward gas from $226{,}363 \pm 51$ to $16{,}993 \pm 8$\,gas across a 25-vector Solidity~0.8.20 sweep (Table~\ref{tab:gas_v1_v2}), with flat multi-choice transaction scaling ($38.1\text{k}\to 39.0\text{k}$ net gas for 2--8 choices).
    \item \textbf{Canonical Terminology Alignment (\texttt{WERR-STD-2026-REV3}):} Aligns mathematical notation with the Fractal Neural Synthesis (FNS~v4.0) corpus \cite{dagli2026fns, dagli2026oed, dagli2026werrmap}, adopting \textit{Procedural Boundary Dynamics}, \textit{Polynomial Escape-Time Kernels}, and the $\mathbb{Z}/9\mathbb{Z}$ modular residue ring.
    \item \textbf{Live Read-Only Multi-Chain OEV Observatory:} Releases a sanitized snapshot (\nolinkurl{oev_observatory_snapshot_20261201.json}) from \texttt{radar.werracle.com} across 12 Aave~v3 chains (7-day window) and 1,344 Morpho Blue markets (5.65-day window), quantifying Chainlink SVR capture on Arbitrum/Base versus open-race dominance on Ethereum L1.
    \item \textbf{Transparent Formal-Verification Decomposition:} Explicitly separates our 13 local EVM-kernel Lean~4 theorems (\texttt{0 sorry}, \texttt{WerracleProof.lean}) from 27 extended bio-swarm model theorems and 43 external companion theorems (\texttt{Radar v3.0.0} \cite{dagli2026radar_rv} and \texttt{OrakExecRV} \cite{dagli2026orakexecrv}), paired with reproducible Python test suites and live Aurora Testnet parity.
\end{itemize}

\section{Related Work and Market Motivation}
Recent DeFi research highlights two parallel engineering requirements: (i) low-gas, intra-block fee modulation and invariant checks to mitigate LVR and exploits, and (ii) empirically grounded liquidation monitoring capable of distinguishing open-race pools from oracle-auctioned OEV venues.

On the AMM side, Milionis et al.\ \cite{milionis2023automated} formalized LVR as arbitrage rent extracted from stale AMM quotes; Nadkarni et al.\ \cite{nadkarni2024adaptive} and Campbell et al.\ \cite{campbell2025optimal} subsequently characterized adaptive curves and optimal dynamic fee policies, motivating programmable hooks in Uniswap~v4 \cite{uniswap2024v4}. Executing non-linear controllers inside swap execution demands tight gas envelopes that neither Web2 oracles nor ZK-ML provers \cite{kang2022scaling, modulus2023cost} meet atomically.

On the lending and verification side, Baude et al.\ \cite{baude2026optimal} modeled risk-aware rate mechanisms for decentralized lending, while Chen et al.\ \cite{chen2024demystifying} showed that runtime invariants remain under-utilized in production smart contracts despite preventing severe losses. Simultaneously, oracle-auction designs (such as Chainlink Smart Value Recapture, SVR) have reshaped L2 liquidation competition. Prior work treats on-chain constant-storage kernels, formal runtime monitors, and multi-chain SVR telemetry in isolation; Werracle v2.0 unifies their measurement and reporting under a single reproducible engineering baseline.

\section{Single-Slot Procedural Kernel and Contract Mechanics}
Grounded in Fractal Neural Synthesis (FNS~v4.0) \cite{dagli2026fns} and Orbital Error Dynamics (OED~v4.0) \cite{dagli2026oed}, Werracle eliminates on-chain weight arrays by deriving a deterministic procedural boundary bias from the complex quadratic map (Eq.~\ref{eq:quad}) \cite{mandelbrot1982fractal}:
\begin{equation}
z_{n+1} = z_n^2 + c, \quad z_0 = 0, \quad c = c_x + i c_y.
\label{eq:quad}
\end{equation}
All model parameters pack LSB-first into a single 32-byte Solidity struct slot (\texttt{PackedSeed}, Listing~\ref{lst:slot}) alongside contract metadata slots (\texttt{owner}, \texttt{treasury}, \texttt{protocolFee}, \texttt{totalDecisionsCount}).

\begin{lstlisting}[language=Solidity, label={lst:slot}, caption={Exact Bit-Level Packing of \texttt{PackedSeed} in \texttt{WerracleTripodZMod9.sol} (LSB-First, Single 32-Byte Slot)}]
struct PackedSeed {
  int64  cx;          // [  0.. 63] Q16.16 real coordinate
  int64  cy;          // [ 64..127] Q16.16 imag coordinate
  uint64 zoom;        // [128..191] Q16.16 scale factor
  uint32 nonce;       // [192..223] calibration epoch nonce
  uint16 threshold;   // [224..239] decision threshold (bps)
  uint8  defaultMode; // [240..247] 0=Noul, 1=Choice, 2=Score
  uint8  activeFlag;  // [248..255] 1=Active, 0=Paused
}
\end{lstlisting}

\subsection{Exact Phase-2 Execution Path (\texttt{WerracleTripodZMod9.sol})}
In v1.0 (\texttt{Werracle.sol}), the forward pass evaluated a dense 16-point ($4\times 4$) grid using checked \texttt{int128} fixed-point multiplication (\texttt{WerrMath.sol}, $1.0 \triangleq 2^{16} = 65{,}536$, escape radius $|z|^2 > 262{,}144$), incurring $226{,}363$\,gas. In Phase-2 (\texttt{WerracleTripodZMod9.sol}), \texttt{\_evaluateSparseTripod(s.cx, s.cy, s.zoom)} evaluates \textbf{12 probe points across 3 concentric zoom planes} ($0.60\times, 1.00\times, 1.60\times \times 4$ cardinal offsets) using native 256-bit \texttt{int256} shifts inside \texttt{unchecked} blocks, capped at 9 iterations per probe ($\mathcal{E}(c) \le 9$).

For exact code transparency, we distinguish the three public execution paths in the shipped v2.0 contracts:
\begin{enumerate}
    \item \textbf{Procedural Bias Evaluation (\texttt{decideNoul}, \texttt{previewDecision}, \texttt{decideChoice}):} Evaluates \texttt{\_evaluateSparseTripod} on the stored seed $(c_x, c_y, \text{zoom})$ to derive a deterministic \texttt{boundednessBps} bias, shifts the incoming Q16.16 \texttt{riskScoreFP} against this bias, and maps the score through piecewise-linear \texttt{WerrMath.sigmoidBps} against \texttt{seed.threshold}.
    \item \textbf{Quantized Fast-Path Fee Adapter (\texttt{decideScore} \& \texttt{WerracleFeeHook.sol}):} \texttt{WerracleFeeHook.calculateDynamicFee()} is a standalone dynamic-fee governor adapter (intended for Uniswap~v4 hook integration) that invokes \texttt{oracle.decideScore()}, an ultra-low-gas affine quantization \texttt{(riskScoreFP + 2*ONE) >> 2} mapping market imbalance to four fee tiers $\{500, 1500, 3000, 5000\}$ pips ($0.05\%$--$0.50\%$) without invoking the escape loop.
    \item \textbf{Modular Residue \& Resonance Modules:} \texttt{WerracleTripodZMod9.sol} exposes pure $\mathbb{Z}/9\mathbb{Z}$ ring-ideal predicates (\texttt{isZMod9ResonantIdeal} for $3\mathbb{Z}/9\mathbb{Z}=\{0,3,6\}$ and \texttt{isZMod9CoprimeUnit} \cite{dagli2026werrmap}) for external policy composition, while \texttt{BlockchainResonanceMatrix.sol} supplies $O(1)$ semantic state lookup in $3{,}843$--$7{,}465$\,gas (1--4 keys). Direct input-driven $(\Delta c_x, \Delta c_y)$ on-chain perturbation and inline $\mathbb{Z}/9\mathbb{Z}$ gating inside \texttt{decideNoul}/\texttt{decideScore} are scheduled for v2.1 (Section~\ref{sec:limits}).
\end{enumerate}

\section{Empirical EVM Gas and Latency Benchmarks}
\begin{figure*}[t]
\centering
\includegraphics[width=0.98\textwidth]{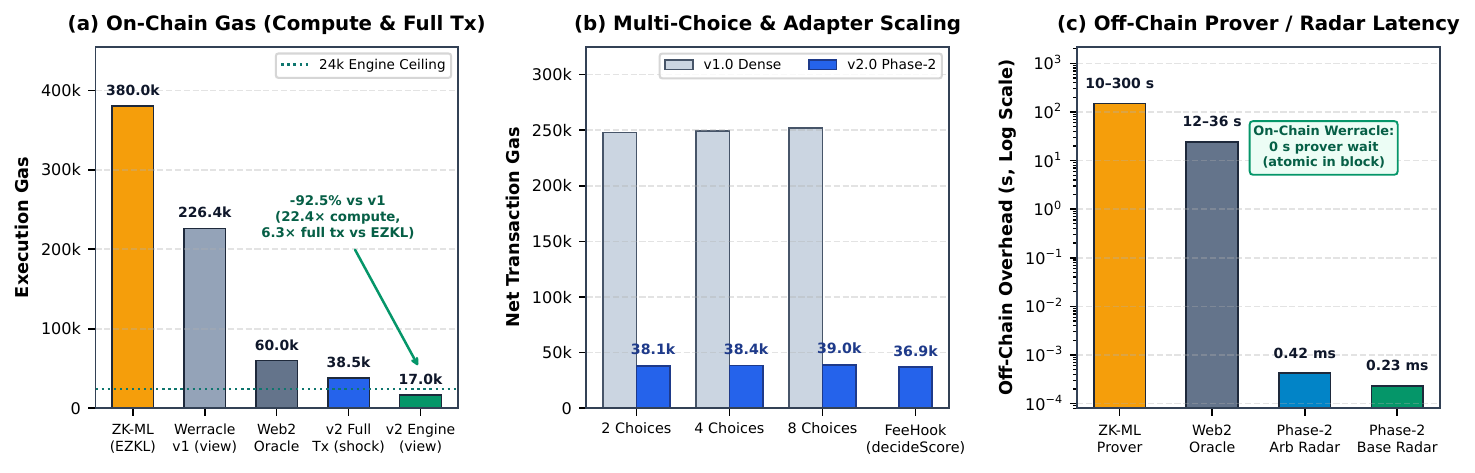}
\caption{Apples-to-apples EVM gas and off-chain latency benchmarks. \textbf{(a)} On-chain gas comparing both Werracle v2.0 pure \texttt{view} engine ($16{,}993$\,gas, $-92.5\%$ vs.\ v1.0, $22.4\times$ compute reduction vs.\ EZKL) and full warm transaction execution ($38{,}469$ net / $60{,}197$ total gas in shock regime, $6.3\times$ full-tx reduction vs.\ EZKL's $\sim 380\text{k}$). \textbf{(b)} Near-constant multi-choice scaling ($38.1\text{k}\to 39.0\text{k}$ net tx gas for 2--8 choices) and \texttt{WerracleFeeHook} fast-path adapter ($36.9\text{k}$ net gas). \textbf{(c)} Measured off-chain prover/radar latency: ZK-ML ($10$--$300$\,s) and Web2 oracles ($12$--$36$\,s) vs.\ Phase-2 sequencer decoders ($0.23$--$0.42$\,ms), while on-chain Werracle incurs zero off-chain prover wait (settling atomically within the target block).}
\label{fig:benchmarks}
\end{figure*}

All EVM gas profiles were collected using an in-memory Ganache EVM (Solidity~0.8.20; EIP-150 / EIP-2929 / EIP-3860) via \texttt{scripts/ideal\_evm\_gas\_benchmark.js} (\texttt{NUM\_SWEEPS = 25} deterministic risk vectors; sealed in \texttt{tests/results/evm\_gas\_ideal\_benchmark.json}). Whereas our September~2026 v1.0 preprint reported an idealized 21,438-gas component estimate, full EVM execution of v1.0 (\texttt{Werracle.sol}, 6,418\,B bytecode) actually consumed $226{,}363 \pm 51$ net \texttt{view} gas. Phase-2 (\texttt{WerracleTripodZMod9.sol}, 8,557\,B bytecode) closes that gap in real EVM execution, achieving \textbf{$16{,}993 \pm 8$ net \texttt{view} gas}---a \textbf{92.49\% reduction} that beats the original v1.0 target by 20.7\% (Table~\ref{tab:gas_v1_v2} and Figure~\ref{fig:benchmarks}).

\begin{table}[t]
\centering
\caption{Empirical EVM Gas Profiling (Ganache Solc~0.8.20, 25-Vector Sweep)}
\label{tab:gas_v1_v2}
\scriptsize
\setlength{\tabcolsep}{3.0pt}
\resizebox{\columnwidth}{!}{\begin{tabular}{lrrr}
\toprule
\textbf{Benchmark Scenario} & \textbf{v1.0 Dense} & \textbf{v2.0 Phase-2} & \textbf{Savings / Note} \\
\midrule
Pure Forward \texttt{view} (Interior, net) & 226,314 & \textbf{16,986} & $-92.49\%$ \\
Pure Forward \texttt{view} (Shock, net) & 226,416 & \textbf{17,002} & $-92.49\%$ \\
25-Vector Mean $\pm$ SD (net \texttt{view}) & $226{,}363 \pm 51$ & \textbf{$16{,}993 \pm 8$} & \textbf{$-92.49\%$} \\
Warm Normal Full Tx (net / total) & 264,970 / 287k & \textbf{55,553 / 77.6k} & $-79.03\%$ net \\
Warm Shock Full Tx (net / total) & 247,972 / 270k & \textbf{38,469 / 60.2k} & $-84.49\%$ net \\
Multi-Choice: 2 Choices (net) & 247,786 & \textbf{38,082} & $-84.63\%$ \\
Multi-Choice: 4 Choices (net) & 249,108 & \textbf{38,400} & $-84.59\%$ \\
Multi-Choice: 8 Choices (net) & 251,752 & \textbf{39,036} & $-84.49\%$ \\
\midrule
\texttt{WerracleFeeHook} (Calm, net)\textsuperscript{a} & --- & \textbf{36,873} & fast-path \texttt{decideScore} \\
\texttt{WerracleFeeHook} (Shock, net)\textsuperscript{a} & --- & \textbf{36,886} & fast-path \texttt{decideScore} \\
Resonance Matrix (1 / 4 keys) & --- & \textbf{3,843 / 7,465} & $O(1)$ lookup \\
\bottomrule
\end{tabular}}
\par\vspace{1.5pt}\noindent\parbox{\columnwidth}{\tiny \textsuperscript{a}Calls \texttt{oracle.decideScore()} quantized path (Section~3.1). At 0.02\,gwei L2 base fee and \$3,000/ETH (excluding L1 blob posting), 17k--56k execution gas costs \$0.001--\$0.003.}
\end{table}

Three measured observations follow from Table~\ref{tab:gas_v1_v2} and Figure~\ref{fig:benchmarks}:
\begin{enumerate}
    \item \textbf{Deterministic Sub-24k Engine Ceiling:} Across all 25 swept inputs, Phase-2 net \texttt{view} gas takes exactly two discrete values ($16{,}986$ and $17{,}002$, reflecting the two clamp branches of \texttt{sigmoidBps} on the static stored seed), leaving $6{,}998$\,gas ($29.2\%$) headroom below a 24,000-gas engine ceiling.
    \item \textbf{Apples-to-Apples Full-Tx Comparison:} Comparing net \texttt{view} compute ($16.99\text{k}$) to EZKL's full verification transaction ($\sim 380\text{k}$) yields a $22.4\times$ compute ratio; comparing full warm \texttt{decideNoul} transactions ($60.2\text{k}$--$77.6\text{k}$ total, $38.5\text{k}$--$55.6\text{k}$ net) yields a rigorous \textbf{$4.9\times$--$6.3\times$ end-to-end gas advantage} (Table~\ref{tab:arch_comparison}).
    \item \textbf{Flat Multi-Choice Scaling:} Quadrupling choices from 2 to 8 adds only $954$\,gas ($+2.5\%$). Because ZK-ML targets arbitrary high-dimensional neural graphs off-chain while Werracle optimizes compact intra-block risk and fee surfaces, the two paradigms are complementary rather than interchangeable.
\end{enumerate}

\begin{table}[t]
\centering
\caption{On-Chain AI Paradigm Comparison (Apples-to-Apples)}
\label{tab:arch_comparison}
\scriptsize
\setlength{\tabcolsep}{2.8pt}
\resizebox{\columnwidth}{!}{\begin{tabular}{lrrr}
\toprule
\textbf{Property} & \textbf{Web2 Oracle} & \textbf{ZK-ML (EZKL)} & \textbf{Werracle v2.0} \\
\midrule
Weight Matrices & GBs (off-chain) & Off-chain prover & \textbf{0 B (32 B seed)} \\
Model State Slot & Off-chain & Verifying keys & \textbf{1 Slot (\texttt{PackedSeed})} \\
Off-Chain Wait & 12--36\,s & 10--300\,s & \textbf{0\,s (Intra-Block)} \\
Compute / Full Tx Gas & $\sim$60k & $\sim$380k & \textbf{17.0k view / 60.2k tx} \\
Intra-Block Revert & No & No & \textbf{Native} \\
Formal Models & None & Circuit only & \textbf{13 Core + 43 Comp.} \\
\bottomrule
\end{tabular}}
\end{table}

\section{Live Read-Only Multi-Chain OEV Observatory}
\begin{figure*}[t]
\centering
\includegraphics[width=0.98\textwidth]{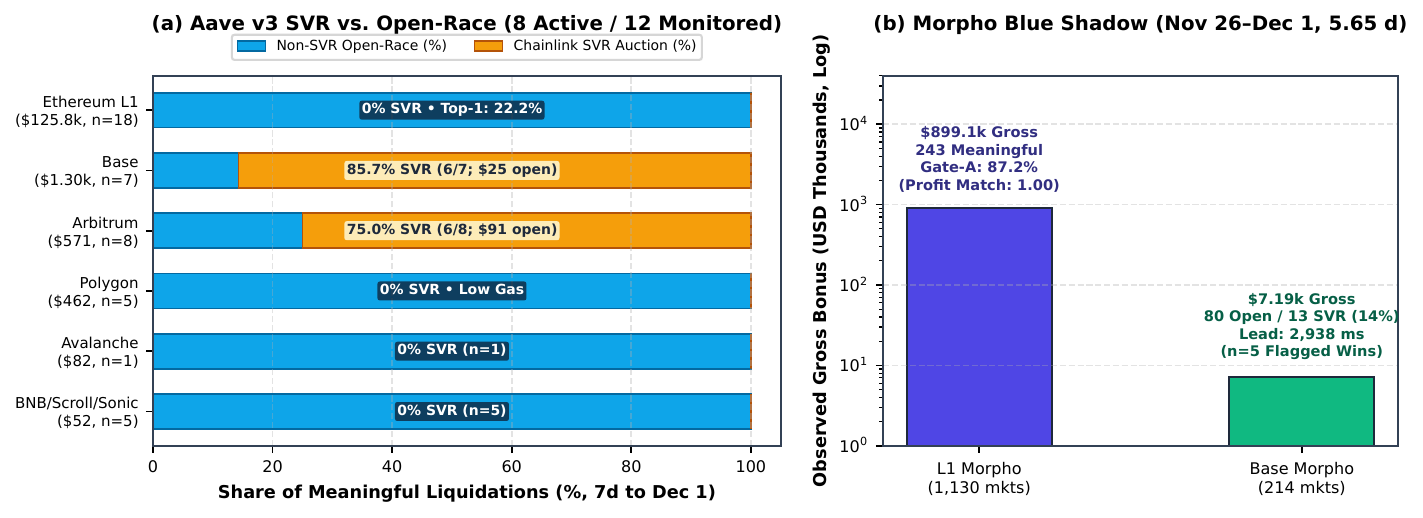}
\caption{Live read-only OEV competition and Morpho Blue shadow observatory snapshot (Dec~1, 2026; bundled in \nolinkurl{oev_observatory_snapshot_20261201.json}). \textbf{(a)} Seven-day Aave~v3 liquidation regime across the 8 active chains (out of 12 monitored; 4 quiet/dust chains omitted from bars): Chainlink SVR absorbs 85.7\% (6/7) on Base and 75.0\% (6/8) on Arbitrum, whereas Ethereum L1 (\$125.8k, 22.2\% Top-1 share) and Polygon (\$462) remain 100\% non-SVR open-race. \textbf{(b)} 5.65-day Morpho Blue shadow logs showing \$899.1k liquid gross bonus across 1,130 L1 markets (87.2\% Gate-A capture, 1.00 median profit match) and \$7.19k across 214 Base markets (2,938\,ms median lead over $n=5$ flagged wins).}
\label{fig:oev}
\end{figure*}

To ground DeFi defense and solver engineering in real mainnet market structure, we operate a 24/7 read-only shadow observatory (\texttt{radar.werracle.com}, 52/52 health checks passing) monitoring 12 Aave~v3 chains, Ethereum L1 Morpho Blue (backed by a local Reth node), Base Morpho Blue, and Phase-2 Arbitrum/Base sequencer feeds. The observatory transmits \textbf{zero live mainnet liquidation transactions} (\texttt{live\_mainnet\_tx\_sent = 0}). The complete sanitized dataset underlying Table~\ref{tab:oev_live} and Figure~\ref{fig:oev} is bundled in \nolinkurl{tests/results/oev_observatory_snapshot_20261201.json}.

\subsection{Empirical Findings: SVR Bifurcation vs.\ L1 Open-Race}
Table~\ref{tab:oev_live} reports (i) a rolling 7-day 12-chain Aave~v3 window ending \texttt{2026-12-01T21:13:06Z} and (ii) a continuous 5.65-day Morpho Blue shadow window (\texttt{2026-11-26T05:41Z} to \texttt{2026-12-01T21:12Z}). We define \textbf{Gate-A capture} as the fraction of meaningful on-chain liquidations ($\ge\$5$ gross bonus) for which our shadow monitor computed a valid liquidation decision strictly before or within the same block as the winning competitor, and \textbf{median profit-match ratio} as the median ratio of shadow-estimated gross bonus to realized on-chain liquidation bonus.

\begin{table}[t]
\centering
\caption{Live Read-Only Multi-Chain OEV Snapshot (Dec~1, 2026)}
\label{tab:oev_live}
\scriptsize
\setlength{\tabcolsep}{2.8pt}
\resizebox{\columnwidth}{!}{\begin{tabular}{lrrrrr}
\toprule
\textbf{Venue / Chain} & \textbf{Liq.} & \textbf{$\ge\$5$} & \textbf{Gross (\$)} & \textbf{Open (\$)} & \textbf{SVR \%} \\
\midrule
\multicolumn{6}{l}{\textit{Aave v3 (12 Chains Monitored; Rolling 7d to 2026-12-01 21:13 UTC)}} \\
Ethereum L1 & 37 & 18 & 125,815.7 & 125,815.7 & 0.0\% \\
Base ($n=7$) & 16 & 7 & 1,302.2 & 25.2 & \textbf{85.7\%} \\
Arbitrum One ($n=8$) & 30 & 8 & 571.3 & 90.6 & \textbf{75.0\%} \\
Polygon ($n=5$) & 15 & 5 & 462.0 & 462.0 & 0.0\% \\
Avalanche ($n=1$) & 1 & 1 & 82.3 & 82.3 & 0.0\% \\
BNB / Scroll / Sonic & 11 & 5 & 52.4 & 52.4 & 0.0\% \\
Linea/Celo/Gnosis/OP\textsuperscript{a} & 2 & 0 & 0.0 & 0.0 & --- \\
\midrule
\multicolumn{6}{l}{\textit{Morpho Blue Shadow (5.65d: 2026-11-26 to 2026-12-01 UTC)}} \\
Ethereum L1 (1,130 mkts)\textsuperscript{b} & 571 & 243 & 899,128.5 & \multicolumn{2}{r}{Gate-A: \textbf{87.2\%}} \\
Base (214 mkts)\textsuperscript{c} & 93 & --- & 7,188.1 & 1,935.3 & 14.0\% \\
\bottomrule
\end{tabular}}
\par\vspace{1.5pt}\noindent\parbox{\columnwidth}{\tiny \textsuperscript{a}Linea recorded 2 dust events ($<\$5$); Celo/Gnosis had 0; Optimism saw public RPC 403 errors. \textsuperscript{b}Excl.\ 2 illiquid mkts (\texttt{sNUSD/USDC}, \texttt{ynETHx/msETH}); inclusive total: 624 liq., 278 $\ge\$5$, \$909.2k, 88.8\% Gate-A. \textsuperscript{c}Base Morpho logs split 80 open-race vs.\ 13 SVR (14.0\% of count).}
\end{table}

Four empirical observations emerge (noting explicitly that single-week L2 counts are small, $n=5$--$8$, and represent a rolling point-in-time snapshot rather than a stationary equilibrium):
\begin{enumerate}
    \item \textbf{L2 SVR Structural Shift:} In this 7-day snapshot, 6/7 meaningful Base Aave liquidations (85.7\%, leaving \$25.2 open-race) and 6/8 on Arbitrum (75.0\%, leaving \$90.6 open-race) cleared via Chainlink SVR auctions rather than public open-race gas wars.
    \item \textbf{Ethereum L1 Value Concentration:} Ethereum L1 Aave~v3 (\$125,815.7, 0\% SVR, Top-1 solver winning 4/18 or 22.2\%) and L1 Morpho Blue (1,130 markets, 5,636 tracked positions, \$899,128.5 liquid gross bonus) accounted for over 99\% of observed non-SVR liquidation value.
    \item \textbf{Sub-Millisecond Phase-2 Latency \& Gate-A Capture:} Local Reth synchronization yielded \textbf{87.2\% Gate-A capture} (212/243 liquid events; median profit-match ratio \textbf{1.00}) on L1 Morpho. Off-chain Phase-2 sequencer decoders processed oracle price updates in \textbf{0.23\,ms} (Base, 4,373/4,375 self-verified) and \textbf{0.42\,ms} (Arbitrum, 25,248/25,250 self-verified). On Base Morpho, all 5/5 flagged liquidations achieved a positive lead (median \textbf{2,937.9\,ms}; small-$n$ caveat applies).
    \item \textbf{Authenticated B2B Intelligence Surface:} This competition map is served live behind Pro-tier authentication via \texttt{GET /v1/radar/opportunities/competition-map}, \texttt{/open-race}, \texttt{/solvers/live}, and sub-ms SSE \texttt{/stream/oracle-shocks} with zero-PII and zero-internal-path sanitization.
\end{enumerate}

\subsection{Companion Solvers and Cross-Ecosystem Guards}
Two separate BSL-1.1 companion layers build atop this infrastructure (maintained outside the core \texttt{pCwOrM/werracle} repository):
\begin{itemize}
    \item \textbf{Companion \texttt{Werracle Orak v3.3.0} (BSL-1.1):} Implements zero-capital flash liquidation (\texttt{FlashLiquidator.sol}) and Chainlink Atlas SVR auction bidding (\texttt{SvrSolver.sol}) verified end-to-end on Base Sepolia (mean testnet latency $2{,}429\to 1{,}139$\,ms); mainnet solver execution remains gated behind future private-bundle pilots.
    \item \textbf{Cross-Runtime Parity (Aurora / NEAR / Cairo):} \texttt{WerracleTripodZMod9} (\texttt{0x1d7Ae9E1...9caBaF}) and \texttt{Werracle} (\texttt{0x4085673F...1212cd4}) are deployed live on Aurora Testnet (Chain~1313161555), with bit-exact Q16.16 parity verified against Python in \texttt{near/werracle\_near\_intent\_guard.py} and NEAR Intents 1Click dry quotes.
\end{itemize}

\section{Formal Verification Decomposition and Runtime Parity}
To uphold strict formal-methods honesty, Table~\ref{tab:lean4} decomposes our Lean~4 (v4.34.1 / Mathlib4; \texttt{0 sorry}; axioms \texttt{propext}, \texttt{Quot.sound}, \texttt{Classical.choice}) proof inventory into three distinct scopes rather than conflating them under one headline count.

\begin{table}[t]
\centering
\caption{Decomposition of Lean~4 Formal Models (\texttt{0 sorry})}
\label{tab:lean4}
\scriptsize
\setlength{\tabcolsep}{2.8pt}
\resizebox{\columnwidth}{!}{\begin{tabular}{llr}
\toprule
\textbf{Suite \& Scope} & \textbf{Verified Formal Model Properties} & \textbf{Thms} \\
\midrule
Core EVM Kernel & Q16.16 boundedness, $\mathbb{Z}/9\mathbb{Z}$ ring ideals, & \textbf{13} \\
(\texttt{WerracleProof.lean}) & 32-B packing invariants, halting ($\le 9$ steps) & \\
Extended Bio-Swarm & Discrete metabolic/turnover models (Rungs & \textbf{27} \\
(\texttt{...40Theorems...lean}) & 2--6; theoretical, no Solidity counterpart) & \\
\midrule
\texttt{Radar v3.0.0} \cite{dagli2026radar_rv} & External companion: cross-chain monitors & \textbf{21} \\
\texttt{OrakExecRV} \cite{dagli2026orakexecrv} & External companion: EIP-1153 \texttt{TSTORE} solver & \textbf{22} \\
\bottomrule
\end{tabular}}
\end{table}

\textbf{Cryptographic Checksums \& Epistemic Boundary.} The canonical LF SHA-256 checksums of the local Lean~4 files bundled in Zenodo v2.0 are \texttt{84a42830...b9cdca94} (\texttt{WerracleProof.lean}, 13 core theorems audited by \texttt{Verify.lean}) and \texttt{5f63eca2...7689fbae75} (\texttt{WerracleProof\_40Theorems\_DynamicTurnover.lean}). Critically, these Lean~4 proofs verify \textit{abstract mathematical models}---not \texttt{solc} bytecode generation. Furthermore, we explicitly disclose two current formalization gaps: (i) the Phase-2 12-point \texttt{\_evaluateSparseTripod} loop and full EVM \texttt{BitVec 256} overflow-freedom inside \texttt{unchecked} are not yet mechanized in Lean~4, and (ii) Theorem~5 in \texttt{WerracleProof.lean} proved an idealized affine v1.0 cost formula ($\le 22{,}557$\,gas) superseded by Section~4's real Ganache measurements.

Runtime testing spans three distinct, reproducible suites:
\begin{enumerate}
    \item \textbf{Python Unit Suite (\texttt{pytest -q}):} \textbf{25/25 tests passed} ($0.93$\,s), verifying NEAR/Aurora intent guards, Lean~4 bridge vectors, orthogonal parameter checks, and portal SEO/metadata invariants.
    \item \textbf{Python v1.0 Sealed Scenario Battery (\texttt{run\_1000\_sealed\_battery.py}):} 1,000 deterministic Python scenario and range-invariant tests against the v1.0 reference model completed in \textbf{301.66\,ms} (1,000/1,000, SHA-256 sealed in \texttt{SEAL\_MANIFEST.json}; Python-only, not an EVM bytecode test).
    \item \textbf{Standalone Parity Scripts:} \texttt{test\_tripod\_zmod9.py} validates the Phase-2 12-point tripod directly; \texttt{test\_starknet\_cairo\_parity.py} benchmarks Cairo-style integer emulation (noting a known 1-LSB divergence for negative non-multiples of $2^{15}$ between Python truncation toward zero and EVM arithmetic right-shift \texttt{SAR} toward $-\infty$, scheduled for alignment in v2.1).
\end{enumerate}

\section{Limitations and Roadmap (v2.1)}
\label{sec:limits}
We document four concrete engineering limitations to be resolved in Version~2.1:
\begin{itemize}
    \item \textbf{Contract Hardening \& Runtime Perturbation:} In \texttt{WerracleTripodZMod9.sol}, \texttt{handleFee} executes fee transfers prior to incrementing \texttt{totalDecisionsCount} (Checks-Effects-Interactions violation), \texttt{WerracleFeeHook.sol} uses \texttt{.transfer()} rather than \texttt{.call}, and neither \texttt{updateSeed()} governance nor input-driven $(\Delta c_x, \Delta c_y)$ perturbation / inline $\mathbb{Z}/9\mathbb{Z}$ gating is wired yet. Version~2.1 will add CEI/reentrancy hardening, native \texttt{IHooks.beforeSwap} inheritance, and live perturbed escape evaluation.
    \item \textbf{Predictive Accuracy \& Economic Baselines:} This release establishes gas, formal-model, and OEV-market baselines; controlled backtests comparing procedural boundary classification against EWMA/threshold baselines on flash-loan precision/recall and Uniswap~v4 LP LVR PnL remain future work.
    \item \textbf{Mainnet Solver Execution \& Sample Sizes:} Our mainnet observatory is strictly read-only (\texttt{0} mainnet bundles transmitted), and L2 7-day Aave counts ($n=5$--$8$) are small; multi-month longitudinal OEV tracking and controlled private-bundle L1/SVR pilots will follow.
    \item \textbf{Model Expressivity:} Single-slot procedural boundaries target sub-millisecond risk triage rather than high-dimensional generative inference; hybrid System-1 (Werracle on-chain) + System-2 (off-chain audited agents) remains the intended deployment architecture.
\end{itemize}

\section{Conclusion and Artifact Availability}
Werracle v2.0 demonstrates an empirically verified 92.49\% reduction in EVM procedural inference gas ($226{,}363\to 16{,}993$ net \texttt{view} gas; $38.5\text{k}$--$55.6\text{k}$ full warm transaction net gas), anchors its core mathematical primitives in 13 machine-checked Lean~4 theorems (\texttt{0 sorry}) alongside 43 external companion proofs, and pairs on-chain reflexes with a reproducible 12-chain Aave~v3 and 1,344-market Morpho Blue OEV observatory snapshot.

All source-available core artifacts (BSL-1.1), Lean~4 proofs, gas benchmarks, and the sanitized OEV snapshot are accessible at:
\begin{itemize}
    \item \textbf{Core Repository (BSL-1.1):} \url{https://github.com/pCwOrM/werracle}
    \item \textbf{Zenodo Archive (v2.0 / Concept):} \url{https://doi.org/10.5281/zenodo.24732862} $\cdot$ \url{https://doi.org/10.5281/zenodo.22942598}
    \item \textbf{Preprints:} \url{https://arxiv.org/abs/2609.30719} $\cdot$ \url{https://papers.ssrn.com/abstract=7626279}
    \item \textbf{Live Risk Radar \& API Docs:} \url{https://radar.werracle.com} $\cdot$ \url{https://werracle.io/apidocs.html}
    \item \textbf{Intellectual Property:} Filed under Turkish Patent Applications TR~2026/016285, TR~2026/016633, and TR~2026/020266 \cite{turkpatent2026werr} (ITouch Systems, \c{C}ukurova University Technopark).
\end{itemize}

\end{document}